\documentclass[prl,aps,floats,twocolumn]{revtex4}

\usepackage[dvips]{graphicx}
\usepackage{amssymb} 
\usepackage{amsmath}
\usepackage{ctable}
\usepackage{slashed}

\begin{document}

\title{Rethinking Connes' approach to the standard model of particle physics via 
non-commutative geometry}

\author{Shane Farnsworth and Latham Boyle} 

\affiliation{Perimeter Institute for Theoretical
Physics, Waterloo, Ontario N2L 2Y5, Canada} 

\date{August 2014}

\begin{abstract}
Connes' non-commutative geometry (NCG) is a generalization of Riemannian geometry that is
particularly apt for expressing the standard model of particle physics coupled to
Einstein gravity.  In a previous paper, we suggested a reformulation of this framework 
that is: (i) simpler and more unified in its axioms, and (ii) allows the Lagrangian for the standard 
model of particle physics (coupled to Einstein gravity) to be specified in a way that is tighter 
and more explanatory than the traditional algorithm based on effective field theory.  Here we 
explain how this same reformulation yields a new perspective on the symmetries of a 
given NCG.  Applying this perspective to the NCG traditionally used to describe the 
standard model we find, instead, an extension of the standard model by an
extra $U(1)_{B-L}$ gauge symmetry, and a single extra complex scalar field $\sigma$,
which is a singlet under $SU(3)_{C}\times SU(2)_{L}\times U(1)_{Y}$, but has $B-L=2$.
This field has cosmological implications, and offers a new
solution to the discrepancy between the observed Higgs mass and the NCG prediction. 
\end{abstract}

\maketitle

\section{Introduction}

Connes' non-commutative geometry (NCG) \cite{Connes:1985, Connes:1994yd} is a 
generalization of Riemannian geometry which also provides a particularly apt framework 
for expressing and geometrically reinterpreting the action for the standard model of particle 
physics, coupled to Einstein gravity \cite{Connes:1990qp, Connes:1995tu, 
Connes:1996gi, Chamseddine:1991qh, Chamseddine:1996zu, Barrett:2006qq, 
Connes:2006qv, Chamseddine:2006ep, Chamseddine:2007hz, Chamseddine:2007ia} (for an 
introduction, see \cite{Chamseddine:2010ps, vandenDungen:2012ky}).  
In a recent paper \cite{Boyle:2014wba}, we suggested a simple 
reformulation of the NCG framework, and pointed out three key advantages of this 
reformulation: (i) it unifies many of the traditional NCG axioms into a single, simpler axiom; 
(ii) it immediately yields a further generalization, from non-commutative to non-associative 
geometry  \cite{Farnsworth:2013nza}; and (iii) it resolves a key problem with the traditional 
NCG construction of the 
standard model, thereby making the NCG construction tighter and more explanatory than 
the traditional one based on effective field theory \cite{BoyleFarnsExpository}.  

Here we report the discovery of three crucial and unexpected consequences of the 
reformulation in \cite{Boyle:2014wba}.  (i) First, it yields a new notion of the natural symmetry 
associated to any non-commutative space, and the action functional that lives on that space.  
(ii) Second, when we work out the realization of this 
symmetry for the non-commutative geometry used to describe the standard model of particle 
physics we find that the usual $SU(3)_{C}\times SU(2)_{L}\times U(1)_{Y}$ 
gauge symmetry is augmented by an extra $U(1)_{B-L}$ factor.  (iii) Third, as a 
consequence of this additional gauge symmetry, we find the standard
model field content must be augmented by the following two fields: a $U(1)_{B-L}$ gauge 
boson $C_{\mu}$, and a single complex scalar field $\sigma$ 
which is an $SU(3)_{C}\times SU(2)_{L}\times U(1)_{Y}$ singlet and has charge $B-L=2$.

The scalar field $\sigma$ has important phenomological implications.  
(i) First, although the traditional NCG
construction of the standard model predicted an incorrect Higgs mass ($m_{h}\approx
170~{\rm GeV}$), several recent works \cite{Chamseddine:2012sw, Devastato:2013oqa, 
Chamseddine:2013sia, Chamseddine:2013rta} have explained that an additional real singlet
scalar field $\sigma$ can resolve this problem, and also restore the stability of the Higgs vacuum.  
Our $\sigma$ field, although somewhat different (since it is complex, and charged under $B-L$), 
solves these same two problems for exactly the same reasons (as may be seen in the
$U(1)_{B-L}$ gauge where $\sigma$ is real).  (ii) Furthermore, precisely this
field content (the standard model, extended by a right-handed neutrino in each generation
of fermions, plus a $U(1)_{B-L}$ gauge boson $C_{\mu}$, and a complex scalar field
$\sigma$ that is a singlet under $SU(3)_{C}\times SU(2)_{L}\times U(1)_{Y}$ but
carries $B-L=2$) has been previously considered \cite{Iso:2009ss, Boyle:2011fq}
because it provides a minimal extension of the standard model 
that can account for several cosmological phenomena that may not be accounted for by the 
standard model alone: namely, the existence of dark matter, the cosmological 
matter-antimatter asymmetry, and the scale invariant spectrum of primordial curvature perturbations.

\section{Symmetries in NCG: a new perspective}

Before turning to NCG, let us recap three key points about ordinary Einstein 
gravity.  (i) A geometry is specified by two pieces of information: a manifold 
${\cal M}$ (which specifies the differential topology), and a metric $g_{\mu\nu}$ 
(which specifies the geometrical information -- distances, angles, curvature).  
(ii) To this geometry, we then assign the Einstein-Hilbert action 
$S=M_{pl}^{2}\int d^{4}x\;g^{1/2} R$.  (iii) The symmetries of this theory are the 
diffeomorphisms of ${\cal M}$: automorphisms of the differential topology that 
leave $S$ invariant.  Now let us explain how to recast and extend these three 
statements in the NCG context.

(i) In NCG, a geometry is traditionally specified by a so-called ``real spectral triple" 
$\{A,H,D,J,\gamma\}$, consisting of a $\ast$-algebra $A$, a hermitian operator $D$, a 
hermitian unitary operator $\gamma$, and an anti-unitary operator $J$, all of which act as
operators on a Hilbert space $H$ and are constrained to satisfy a list of axioms that
relate them to one another (see Secs.~1 and 2 in Ref.~\cite{vandenDungen:2012ky} for an 
introduction).  Ref.~\cite{Boyle:2014wba} shows that these various elements 
naturally fuse to form a new algebra $B$.  This reformulation in terms of $B$ is more unified (in the sense that many 
traditional NCG axioms then follow from the single requirement that $B$ is an associative 
$\ast$-algebra) and more general (in the sense that the new formalism continues to 
cohere even when the underlying algebras, $A$ and $B$, are taken to be non-associative).
In this generalization of ordinary differential geometry, we can think of $B$ as carrying 
the information about the differential topology, while $D$ carries the information about
the metric.

(ii) To this geometry, we assign the so-called spectral action $S={\rm Tr}[f(D/\Lambda)]
+\langle h|D|h\rangle$, where $\Lambda$ is a constant (with units of energy), 
$f(x)$ is a function from $\mathbb{R}\to\mathbb{R}$ that vanishes sufficiently rapidly for 
$x\gg1$, and $h$ is an arbitrary element of $H$ (see Sec.~3 in \cite{vandenDungen:2012ky} for 
details)\footnote{Here we have written the fermionic action as $\langle h|D|h\rangle$ for
simplicity, but the precise form depends on the KO dimension -- see 
\cite{Chamseddine:2006ep} and Sec. 16.2 in \cite{ConnesMarcolli}.}.

(iii) Now we want to characterize the symmetries of this theory, and how those symmetries
act on the fields.  Here we come to this paper's new contribution: the 
reformulation in terms of $B$ sheds new light on this issue.  

To begin, we briefly recap how $B$ is constructed (for details, see \cite{Boyle:2014wba}).  
First, from $A$, we generate $\Omega A=\Omega^{0}\!A\oplus\Omega^{1}\!A\oplus\Omega^{2}\!A
\oplus\ldots$, the differential graded $\ast$-algebra of forms over $A$.~Then, from $\Omega A$ and $H$, we construct the $\ast$-algebra
$B=\Omega A\oplus H$ by equipping its elements $b=\omega+h$
and $b'=\omega'+h'$ with the product
\begin{subequations}
  \begin{equation}
    \label{B_product}
    bb'=(\omega+h)(\omega'+h')=\omega\omega'+\omega h'+h\omega'
  \end{equation}
  and the anti-automorphism
  \begin{equation}
    \label{B_star}
    b^{\ast}=\omega^{\ast}+Jh,
  \end{equation}
\end{subequations}
where $\omega \omega'\in\Omega A$ is the product inherited from $\Omega A$, while 
$\omega h'\in H$ and $h\omega'\in H$ are bilinear products that define the left-action 
and right-action of $\Omega A$ on $H$.  Notice that $B$, like $\Omega A$, is also a 
{\it graded} algebra, where we may think of $H$ as $\Omega_{}^{\infty}\!A$.  (The interesting 
consequences of this simple observation will be explained in \cite{InfinityForms}.)

Now we can easily characterize the relevant symmetries: as in the Riemannian case, 
they are the automorphisms of the differential topology that leave the action invariant.  
In the NCG context, the automorphisms of the differential topology are simply
the automorphisms of the graded $\ast$-algebra $B$ -- {\it i.e.}\ the invertible linear 
transformations $\alpha:B\to B$ that also preserve the grading, product
and $\ast$-operation on $B$:
\begin{subequations}
  \label{alpha_conditions}
  \begin{eqnarray}
    \label{alpha_grading}
    &\alpha=\alpha_{0}\oplus\alpha_{1}\oplus\alpha_{2}\oplus\ldots\quad
    (\alpha_{n}:\Omega^{n}\!A\to\Omega^{n}\!A),& \\
    \label{alpha_product}
    &\alpha(bb')=\alpha(b)\alpha(b'),& \\
    \label{alpha_star}
    &\alpha(b^{\ast})=\alpha(b)^{\ast}.& 
   \end{eqnarray}
  The fact that diffeomorphisms are either orientation-preserving or orientation-reversing 
  translates to the NCG condition that $\alpha_{\infty}:\Omega^{\infty}\!A\!\to\!\Omega^{\infty}\!A$
  ({\it i.e.}\ $\alpha_{\infty}^{}:H\!\to\! H$) either commutes or 
  anti-commutes with the orientation $\gamma$
  \begin{equation}
    \label{alpha_orientation}
    \alpha_{\infty}^{}\gamma\alpha_{\infty}^{-1}=\pm\gamma,
  \end{equation}
  while the condition that the spectral action is invariant translates to the 
  requirement that $\alpha_{\infty}$ is unitary 
  \begin{equation}
    \label{alpha_unitary}
    \alpha_{\infty}^{\dagger}=\alpha_{\infty}^{-1}.
  \end{equation}
\end{subequations}
Next we translate our conditions (\ref{alpha_grading}, \ref{alpha_product}, \ref{alpha_star}, 
\ref{alpha_orientation}, \ref{alpha_unitary}) on the automorphism $\alpha={\rm e}^{\delta}$ into conditions
on its infinitessimal generator, the derivation $\delta$:
\begin{subequations}
  \label{delta_conditions}
  \begin{eqnarray}
    \label{delta_grading} 
    &\delta=\delta_{0}\oplus\delta_{1}\oplus\delta_{2}\oplus\ldots
    \quad(\delta_{n}:\Omega^{n}\!A\to\Omega^{n}\!A),& \\
    \label{delta_product}
    &\delta(bb')=\delta(b)b'+b\delta(b'),& \\
    \label{delta_star}
    &\delta(b^{\ast})=\delta(b)^{\ast},& \\
    \label{delta_orientation}
    &[\delta_{\infty}^{},\gamma]=0,& \\
    \label{delta_unitary}
    &\delta_{\infty}^{\dagger}=-\delta_{\infty}^{}.&
  \end{eqnarray}
\end{subequations}
So far, we have characterized the {\it classical} symmetries associated
to a given NCG; these will generate the symmetries of the corresponding 
classical gauge theory obtained from the spectral action.  In order for these
gauge symmetries to remain consistent at the quantum level, they must also be 
anomaly free.  If $\{\delta_{\infty}^{\alpha}\}$ denotes a basis for the space 
of all operators $\delta_{\infty}$ obtained by satisfying the restrictions 
(\ref{delta_conditions}), then anomaly freedom corresponds to the additional
constraint
\begin{equation}
  \label{anomaly_constraint}
  {\rm Tr}[\gamma\,\delta_{\infty}^{\alpha}\{\delta_{\infty}^{\beta},\delta_{\infty}^{\gamma}\}]=0
\end{equation}
for any basis elements $\delta_{\infty}^{\alpha}$, $\delta_{\infty}^{\beta}$ and $\delta_{\infty}^{\gamma}$
-- see Eq. (20.81) in Ref.~\cite{PeskinSchroeder}.  In contrast to the classical constraints 
(\ref{delta_conditions}), we do not know if the quantum constraint (\ref{anomaly_constraint}) has a 
more fundamental geometric reinterpretation in our formalism.

\section{Application to the Standard Model: \newline
I. Symmetries and Fermion Charges}

Let us apply this formalism to the NCG traditionally
used to describe the standard model of particle physics
(coupled to Einstein gravity).  The detailed 
spectral triple $\{A,H,D,J,\gamma\}$ (which may be fused
into an algebra $B$) is reviewed pedagogically 
in \cite{vandenDungen:2012ky}.  It is the product of two 
triples: the canonical Riemannian triple 
$\{A_{c},H_{c},D_{c},J_{c},\gamma_{c}\}$ (which may be
fused into an algebra $B_{c}$), and the finite 
triple $\{A_{F},H_{F},D_{F},J_{F},\gamma_{F}\}$ (which may 
be fused into an algebra $B_{F}$).  The derivations $\delta$ of
$B$ will involve two types of contributions: those coming 
from the derivations $\delta_{c}$ of $B_{c}$, and those coming
from the derivations $\delta_{F}$ of $B_{F}$.  Here 
we focus on the derivations $\delta_{F}$ and their implications.
In a subsequent paper \cite{FarnsBoyleGravity}, we treat the 
derivations $\delta_{c}$ and their implications.

See \cite{Boyle:2014wba} for a succinct introduction to the finite spectral 
triple $\{A_{F},H_{F},D_{F},J_{F},\gamma_{F}\}$, and the corresponding associative
graded $\ast$-algebra $B_{F}=\Omega^{0}\!A_{F}
\oplus\Omega^{1}\!A_{F}\oplus\ldots$ We will stick to the notation used there.

We would like to find all symmetries of this geometry.  As explained in the previous 
section, this is done by finding all derivations $\delta:B_{F}\to B_{F}$ satisfying 
Eqs.~(\ref{delta_conditions}, \ref{anomaly_constraint}).  First focus on the subalgebra 
$A_{F}=\Omega^{0}\!A_{F}$ and its derivations $\delta_{0}:\Omega^{0}\! A_{F}\to\Omega^{0}\!
A_{F}$: since $\Omega^{0}\!A_{F}$ is a finite-dimensional semi-simple associative 
$\ast$-algebra, its general derivation is given by $\delta_{0}=L_{a}-R_{a}$ \cite{Schafer}, 
where $a=-a^{\ast}$ is any anti-hermitian element of $A_{F}$, and $L_{a}$ and $R_{a}$
denote, respectively, the left-action and right-action of $a$: $L_{a}\omega=a\omega$,
$R_{a}\omega=\omega a$.  We can extend this to a 
derivation on $B_{F}$ by taking $\delta_{n}=L_{a}-R_{a}$ (for all $n=0,1,2,\ldots$).
To display these derivations more explicitly, let us denote an element
of the algebra $A_{F}=\mathbb{C}\oplus\mathbb{H}\oplus M_{3}(\mathbb{C})$ by
$a=(\lambda,q,m)$ where $\lambda\in\mathbb{C}$ is a complex number,
$q\in\mathbb{H}$ is a quaternion, and $m\in M_{3}(\mathbb{C})$ is a $3\times 3$
complex matrix.  We can split the anti-hermitian elements of $A_{F}$ into 3 pieces: 
namely (i) $a_{1}=(\lambda,0,\mu\mathbb{I}_{3})$ where $\lambda\in\mathbb{C}$ 
and $\mu\in\mathbb{C}$ are pure imaginary and $\mathbb{I}_{3}$ is the $3\times3$ 
identity matrix, (ii) $a_{2}=(0,q,0)$ where $q$ is a general anti-hermitian $2\times2$ 
matrix, and (iii) $a_{3}=(0,0,m)$ where $m$ is a general traceless anti-hermitian 
$3\times3$ matrix.  Demanding that the corresponding symmetry generators $\delta_{\infty}^{(i)}
=L_{a_{i}}-R_{a_{i}}$ are anomaly free (\ref{anomaly_constraint}) yields the
additional restriction $\mu=-\lambda/3$.  The $\delta_{\infty}^{(i)}$ are block 
diagonal; if, following \cite{Boyle:2014wba}, we label the subspaces of $H_{F}$ as 
$\{L_{R},Q_{R},L_{L},Q_{L},\bar{L}_{R},\bar{Q}_{R},\bar{L}_{L},\bar{Q}_{L}\}$, 
the blocks are 
\begin{subequations}
  \begin{eqnarray}
    \label{delta1}
    \delta_{\infty}^{(1)}\!&\!=\!&\!\{
    y^{(l)}_{R},y^{(q)}_{R}\otimes\mathbb{I}_{3},y^{(l)}_{L},y^{(q)}_{L}\otimes\mathbb{I}_{3},
    \nonumber \\
    \!&\!\!&\!\;\;\bar{y}^{(l)}_{R},\bar{y}^{(q)}_{R}\otimes\mathbb{I}_{3},\bar{y}^{(l)}_{L},\bar{y}^{(q)}_{L}\otimes\mathbb{I}_{3}\} \\
    \label{delta2}
    \delta_{\infty}^{(2)}\!&\!=\!&\!
    \{0,0,q,q\otimes\mathbb{I}_{3},0,0,\bar{q},\bar{q}\otimes\mathbb{I}_{3}\} \\
    \label{delta3}
    \delta_{\infty}^{(3)}\!&\!=\!&\!
    \{0,\mathbb{I}_{2}\!\otimes\!m,0,\mathbb{I}_{2}\!\otimes\!m,
    0,\mathbb{I}_{2}\!\otimes\!\bar{m},0,\mathbb{I}_{2}\!\otimes\!\bar{m}\}
  \end{eqnarray}
\end{subequations}  
where 
\begin{eqnarray}
  y_{L}^{(l\;\!)}=2\lambda\left(\begin{array}{cc} -\frac{1}{2} & \;\;0\;\; \\ \;\;0\;\; & -\frac{1}{2} \end{array}\right),
  &&y_{R}^{(l\;\!)}=2\lambda\left(\begin{array}{cc} \;\;0\;\; & \;\;0\;\; \\  \;\;0\;\; & -1 \end{array}\right), \nonumber\\
  y_{L}^{(q)}=2\lambda\left(\begin{array}{cc} +\frac{1}{6} & \;\;0\;\; \\ \;\;0\;\; & +\frac{1}{6} \end{array}\right),
  &&y_{R}^{(q)}=2\lambda\left(\begin{array}{cc} +\frac{2}{3} & \;\;0\;\; \\  \;\;0\;\; & -\frac{1}{3} \end{array}\right).
\end{eqnarray}   
In other words, from the derivations $\delta_{n}=L_{a}-R_{a}$ we precisely obtain the generators
$\delta_{\infty}^{(1)}$, $\delta_{\infty}^{(2)}$ and $\delta_{\infty}^{(3)}$ of the familiar standard
model gauge group $U(1)_{Y}\times SU(2)_{L}\times SU(3)_{C}$, with the right- and left-handed
leptons and quarks transforming in their familiar representations.

But $\delta_{n}=L_{a}-R_{a}$ is not the most general possible extension of 
$\delta_{0}=L_{a}-R_{a}$; more generally, we can take $\delta_{n}=L_{a}-R_{a}
+T_{n}$, where the linear operators $T_{n}$ may be non-zero for $n\geq1$, as 
long as they satisfy
\begin{subequations}
  \label{T_conditions}
  \begin{eqnarray}
     \label{T_grading}
     &T_{m}:\Omega^{m}\!A_{F}\to\Omega^{m}\!A_{F},& \\
     &T_{m+n}(\omega_{m}\omega_{n})=(T_{m}\omega_{m})\omega_{n}
     +\omega_{m}(T_{n}\omega_{n}),& \label{T_product} \\
     \label{T_star}
     &T_{m}(\omega_{m}^{\ast})=(T_{m}\omega_{m})^{\ast},& \\
     \label{T_orientation}
     &[T_{\infty},\gamma]=0,& \\
     \label{T_unitarity}
     &T_{\infty}^{\dagger}=-T_{\infty},&
  \end{eqnarray}
\end{subequations}
for any $\omega_{m}\in\Omega^{m}\!A_{F}$ and $\omega_{n}\in\Omega^{n}\!A_{F}$.
If we specialize to the case $(m=0, n=\infty)$ or $(m=\infty, n=0)$ \footnote{These are
the two cases that yield a non-trivial constraint on $T_{\infty}$, while the remaining 
cases end up constraining the other $T_{m}$ (with finite $m$).  As we shall see in the
next section, since $D$ (the object that ultimately appears in the action) is an 
operator on $\Omega^{\infty}\!A=H$, it is only the $\delta_{\infty}$ part of the derivation
$\delta$ that contributes to $D$'s fluctuations, while the $\delta_{m}$ with finite $m$
do not contribute.  Thus, for our purposes (determining the fluctuations 
of $D$, and hence inferring the physical particle content and the symmetries of 
the action) it suffices to determine $T_{\infty}$.}, and use the fact
that $T_{0}=0$, Eqs.~(\ref{T_product}, \ref{T_star}) become
\begin{align}
  \label{T_product_prime}
  [T_{\infty},L_{a}]=[T_{\infty},R_{a}]=0, \tag{\ref{T_product}$'$}\\
  \label{T_star_prime}  
  [T_{\infty},J_{F}]=0,\qquad\;\; \tag{\ref{T_star}$'$}
\end{align}
where $L_{a}$ and $R_{a}$ denote the left or right action of any element $a\in A_{F}$
on $H$.  It is straightforward to check that the most general matrix $\delta_{\infty}=T_{\infty}$ which 
satisfies the constraints (\ref{T_conditions})
along with the anomaly constraint (\ref{anomaly_constraint}) is diagonal, and given by a general
linear combination of the hypercharge generator $\delta_{\infty}^{(1)}$ (\ref{delta1}) and
another generator  
\begin{eqnarray}
  \delta_{\infty}^{(1)'}\!&\!=\!&\!\{
    x_{l},x_{q}\!\otimes\!\mathbb{I}_{3},x_{l},x_{q}\!\otimes\!\mathbb{I}_{3},
    \bar{x}_{l},\bar{x}_{q}\!\otimes\!\mathbb{I}_{3},\bar{x}_{l},\bar{x}_{q}\!\otimes\!\mathbb{I}_{3}\}, 
\end{eqnarray}
where
\begin{equation}
  x_{l}=i\left(\begin{array}{cc} -1 & 0 \\ \;\;0\;\; & -1 \end{array}\right),\quad
  x_{q}=i\left(\begin{array}{cc} \frac{1}{3} & \;\;0\;\; \\  \;\;0\;\; & \frac{1}{3} \end{array}\right).
\end{equation}   
This $\delta_{\infty}^{(1)'}$ generates an extra $U(1)$ symmetry: it is nothing but $U(1)_{B-L}$ (baryon minus lepton number).  The full gauge symmetry associated to the algebra $B_{F}$ is thus 
$SU(3)_{C}\times SU(2)_{L}\times U(1)_{Y}\times U(1)_{B-L}$.

\section{Application to the Standard Model: \newline
II. Bosonic fields and charges}

Under an automorphism $\alpha:B\to B$, the Dirac operator $D:H\!\to\!H$ must transform covariantly: 
$D\!\to\! D'\!=\!\alpha_{\infty}D \alpha_{\infty}^{-1}\approx D-[D,\delta_{\infty}]$.  As in ordinary
gauge theory, by inspecting the fluctuation term $[D,\delta_{\infty}]$, we can read off
the ``connection" terms which must be added to $D$ in order to make it covariant.  In this paper,
$\delta_{\infty}$ means $\delta_{\infty,F}(x)$ (as explained above, this paper focuses 
on the derivations $\delta_{F}$ of $B_{F}$ obtained in the previous section, while 
the derivations $\delta_{c}$ of $B_{c}$, which relate to local lorentz invariance, are treated 
in a subsequent paper \cite{FarnsBoyleGravity}).  The Dirac operator $D$ on the product space is the 
sum of two terms, $D=D_{c}\otimes\mathbb{I}_{F}+\gamma_{c}\otimes D_{F}$, where 
$D_{c}=\gamma^{\mu}\nabla_{\mu}$ is the ordinary curved space Dirac operator, 
while $D_{F}$ is a finite dimensional Hermitian matrix (see \cite{vandenDungen:2012ky});
thus its fluctuation has two terms as well:
\begin{equation}
  \label{fluctuation}
  [D,\delta_{\infty}]=[D_{c}\otimes\mathbb{I}_{F},\delta_{\infty}]
  +[\gamma_{c}\otimes D_{F},\delta_{\infty}].
\end{equation}
Although it may be expressed in unfamiliar notation, the first ($D_{c}$)  term on the right-hand side of 
(\ref{fluctuation}) is nothing but the familiar term that, in ordinary gauge theory, forces one to introduce 
a gauge field $A_{\mu}^{a}$ corresponding to each generator $t^{a}$ of the gauge group [in
this case, $SU(3)_{C}\times SU(2)_{L}\times U(1)_{Y}\times U(1)_{B-L}$] in order to make
the derivatives transform covariantly (see \cite{vandenDungen:2012ky} for more details).  In an analogous
way, the second ($D_{F}$) term on the right-hand side of (\ref{fluctuation}) forces us to add extra fields;
but, whereas the first term involves the regular curved space Dirac operator 
$D_{c}=\gamma^{\mu}\nabla_{\mu}$, and thus induces fields $\gamma^{\mu}A_{\mu}$ with a spacetime 
index $\mu$, the second term involves the finite matrix $D_{F}$, with no spacetime index, and thus 
induces fields with no spacetime index -- {\it i.e.}\ scalar fields (again, see \cite{vandenDungen:2012ky}).  
This is one of the most important advantages of Connes' approach: the gauge fields
and scalar fields and their properties emerge hand in hand, from a single formula, as an inevitable 
consequence of covariance (in constrast to the standard approach, where the gauge fields and
their properties emerge this way, but the scalar fields and their properties do not, and must
instead just be added to the theory by hand).  Let us now compute the $D_{F}$ term in (\ref{fluctuation}) 
and inspect the result.

As explained in \cite{Boyle:2014wba}, there are only four matrices $D_{F}$ compatible with the 
associative algebra $B_{F}$.  The one which is relevant to describing the standard model is given (in the basis 
$\{L_{R},Q_{R},L_{L},Q_{L},\overline{L}_{R},\overline{Q}_{R},\overline{L}_{L},\overline{Q}_{L}\}$) by
\begin{align}
D_F =\begin{pmatrix}
0&0&Y_l^\dagger&0&m^\dagger&0&0&0\\
0&0&0&Y_q^\dagger&0&0&0&0\\
Y_l&0&0&0&0&0&0&0\\
0&Y_q&0&0&0&0&0&0\\
m&0&0&0&0&0&Y_l^T&0\\
0&0&0&0&0&0&0&Y_q^T \\
0&0&0&0&\overline{Y}_l&0&0&0\\
0&0&0&0&0&\overline{Y}_q&0&0
\end{pmatrix},\label{inhomo} 
\end{align}
where $Y_l$ and $Y_q$ are arbitrary $2\times 2$ matrices that act on the doublet indices 
in the lepton and quark sectors, respectively, $m={\rm diag}\{M,0\}$ is $2\times2$ diagonal,
and for brevity we have written $Y_{q}$ in place of $Y_{q}\otimes\mathbb{I}_{3}$.
Thus, if we calculate the fluctuation $D\to D'\approx D-[\gamma_{c}\otimes D_{F},
\delta_{\infty}]$, where $\delta_{\infty}=\delta_{\infty}^{(3)}(x)+\delta_{\infty}^{(2)}(x)
+\delta_{\infty}^{(1)}(x)+\alpha(x)\delta_{\infty}^{(1)'}$, we find $Y_{l}$, $Y_{q}$ and $m$ 
transform as
\begin{subequations}
\begin{align}
Y_l' &=Y_{l\,} - Y_{l\,} q_\lambda(x) + q(x) Y_l\\
Y_q' &=Y_q - Y_q q_\lambda(x) + q(x)Y_q\\
m'&= m + 2i\alpha(x) m\label{sigmafluc} 
\end{align}\label{inhomoeqs}
\end{subequations}
where $q_{\lambda}(x)={\rm diag}\{\lambda(x),\overline{\lambda}(x)\}$.  From this, we read off that, 
to make $D$ covariant, $Y_{l}$ and $Y_{q}$ and $m$ must be promoted to fields
\begin{align}
Y_l \rightarrow \begin{pmatrix}
Y_\nu \phi_1& Y_e\psi_{1} \\
Y_\nu \phi_2 & Y_e\psi_{2}
\end{pmatrix},\quad Y_q \rightarrow \begin{pmatrix}
Y_u \phi_1 & Y_d\psi_{1} \\
Y_u \phi_2 & Y_d\psi_{2}
\end{pmatrix},
\end{align}
and $m\to {\rm diag}\{\sigma,0\}$, where $\{\varphi_{1}(x),\varphi_{2}(x)\}$ and 
$\{\psi_{1}(x),\psi_{2}(x)\}$ are scalar fields that transform as $SU(2)_{L}$ doublets, with 
hypercharge $y=+1/2$ and $y=-1/2$, respectively, and $\sigma(x)$ transforms with charge 
$+2$ under $U(1)_{B-L}$, but is a singlet under $SU(3)_{C}\times SU(2)_{L}\times U(1)_{Y}$.  
Finally, as explained in Ref.~\cite{Boyle:2014wba}, one can 
choose the embedding of $\mathbb{C}$ in $\mathbb{H}$ so that $\{\psi_{1},\psi_{2}\}=
\{-\bar{\varphi}_{2},\bar{\varphi}_{1}\}$; in this way, instead of obtaining a 2-higgs doublet model,
one obtains a single higgs doublet $\{\varphi_{1},\varphi_{2}\}$.  

\section{Discussion}

In our previous paper \cite{Boyle:2014wba}, we found a reformulation of NCG that simplified
and unified the mathematical axioms while, at the same time, resolving a problem with the 
NCG construction of the standard model Lagrangian, by precisely eliminating 7 terms which 
had previously been problematic.  In this paper, we show that this same reformulation leads
to a new perspective on the gauge symmetries associated to a given NCG, uncovering some 
that were previously missed.  In particular, when we apply our formalism to the NCG traditionally 
used to describe the standard model of particle physics, we find a new $U(1)_{B-L}$ gauge 
symmetry (and, correspondingly, a new $B-L$ gauge boson).  This, in turn, implies the existence
of a new complex Higgs field $\sigma$ that is a singlet under 
$SU(3)_{C}\!\times\!SU(2)_{L}\!\times\!U(1)_{Y}$ but transforms with charge $+2$ under $U(1)_{B-L}$,
allowing it to form a majorana-like Yukawa coupling $\sigma\nu_{R}^{}\nu_{R}^{}$ with two right-handed neutrinos (so that, if
it obtains a large VEV, it induces see-saw masses for the neutrinos).  It is striking, on the 
one hand, that this precise extension of the standard model has been previously considered in the 
literature \cite{Iso:2009ss, Boyle:2011fq} on the basis of its cosmological advantages; and, on 
the other hand, that the new field $\sigma$ can resolve a previous discrepancy between the 
observed Higgs mass and the NCG prediction \cite{Chamseddine:2012sw, Devastato:2013oqa, 
Chamseddine:2013sia, Chamseddine:2013rta}.  Note that in the previous works 
\cite{Chamseddine:2012sw, Devastato:2013oqa, Chamseddine:2013sia, Chamseddine:2013rta}
which introduced the $\sigma$ field for this purpose, it was a real field, and a gauge singlet.  By contrast, from
the perspective presented here, the fact that $\sigma$ is complex, and transforms under 
$U(1)_{B-L}$, is the key to its existence: had it been real, it would not have been induced 
by the covariance argument of the previous section.  

It is important to carefully reconsider the phenomenological and cosmological implications of the
standard model extension which we have landed on here, especially in light of the 
extra constraints imposed by the spectral action.  This is an exciting topic for future work.

We would like to thank J. Barrett, L. Dabrowski, G. Landi, F. Lizzi, M. Marcolli and P. Martinetti for valuable discussions.
Research at the Perimeter Institute is supported by the Government of Canada through Industry 
Canada and by the Province of Ontario through the Ministry of Research \& Innovation. LB also
acknowledges support from an NSERC Discovery Grant.

\end{document}